\documentclass[aps,prb,showpacs,superscriptaddress,amsmath,twocolumn]{revtex4}

\usepackage{amsmath}
\usepackage{amssymb}
\usepackage{graphicx}

\begin{document}


\title{Trigonal warping  and anisotropic band splitting in monolayer graphene due to Rashba
spin-orbit coupling}

\author{P. Rakyta}
\affiliation{Department of Physics of Complex Systems,
E{\"o}tv{\"o}s University,
H-1117 Budapest, P\'azm\'any P{\'e}ter s{\'e}t\'any 1/A, Hungary
}

\author{A. Korm\'anyos}
\thanks{e-mail: kormanyos@lancaster.ac.uk}
\affiliation{
Department of Physics, Lancaster University, Lancaster, LA1 4YB, UK}

\author{J. Cserti}
\affiliation{Department of Physics of Complex Systems,
E{\"o}tv{\"o}s University,
H-1117 Budapest, P\'azm\'any P{\'e}ter s{\'e}t\'any 1/A, Hungary
}


\begin{abstract}
We study  the electronic band structure of monolayer graphene when Rashba spin-orbit coupling is
present. We show that if the  Rashba spin-orbit coupling is stronger than
the intrinsic spin-orbit coupling, the low energy bands undergo
trigonal-warping deformation and that for energies smaller than the Lifshitz energy,
the Fermi circle breaks up into separate parts. The effect is very similar to what
happens in bilayer graphene at low energies.  We discuss the possible experimental
implications, such as threefold increase of the minimal conductivity for low electron densities,
anisotropic, wavenumber dependent  spin-splitting of the bands and the spin polarization structure.

\end{abstract}


\maketitle


Following the  pioneering work of Kane and Mele\cite{kane-mele} where
a novel type of quantum spin-Hall effect was predicted,
there has been a significant interest in studying the spin-orbit (SO) coupling
in graphene. Generally, the SO interaction in graphene can have intrinsic
or extrinsic origins.
Density functional and other calculations indicate\cite{ref:ISO-calcs,ref:gmitra,ref:huertas-1}
that the intrinsic spin orbit (ISO)  coupling is in the range  of
 $1-50 \, \mu{\rm eV}$ and therefore it could only be probed at very low temperatures.
The SO interaction can also be induced externally, by impurities\cite{PhysRevLett.103.026804},
by external electric field perpendicular to the graphene sheet or by
electrostatic interaction with the  substrate\cite{PhysRevLett.101.157601}.
The latter two mechanism  can introduce Rashba type spin-orbit (RSO) interaction,
which will be in the focus of our interest in this paper.
Many experimental implications of both intrinsic and extrinsic SO coupling in graphene
have also been addressed in the
literature\cite{kane-mele,rashbacikk,ref:kuemmeth,ref:ingenhoven,ref:ertler,ref:huertas-2,PhysRevLett.103.026804}.
The recent  experiment of Varykhalov \emph{et al}\cite{PhysRevLett.101.157601} have
shown that it was possible to fabricate a quasi freestanding graphene on  Ni(111)
surface with a single layer of intercalated gold atoms between the graphene layer and
the Ni surface. A spin-orbit interaction induced  band splitting  of $\approx 13\, {\rm meV}$
was measured, which is two-three orders of magnitude larger than the predictions for
ISO\cite{ref:huertas-1,ref:ISO-calcs,ref:gmitra}. The spin-orbit coupling was identified as RSO and the
large splitting was attributed to the high nuclear charge of the intercalating gold atoms.
Very recently, Gierz \emph{et al}  have also observed a large and anisotropic spin splitting in
 graphene samples on SiC substrate.

The microscopic Hamiltonian describing the RSO coupling in graphene, shown in Eq.~(\ref{eq:H_RSO}),
 was introduced by  Kane and Mele in Ref.~\onlinecite{ref:kane-mele-2} and its properties were subsequently
studied in e.g. Refs.~\onlinecite{rashbacikk,ref:kuemmeth,PhysRevB.79.165442,PhysRevB.80.115420}.
Here we report on an interesting feature of the low energy spectrum of this Hamiltonian which
has attracted  little attention so far but which can be important in systems where
RSO coupling is much stronger than ISO.  Examples for such systems appear to be the ones studied
in the experiments of  Varykhalov \emph{et al}\cite{PhysRevLett.101.157601}
and Gierz \emph{et al}\cite{ref:gierz}.
We show that close to the $\mathbf{K}$ and
$\mathbf{K}'$ points of the graphene's Brillouin zone the  Hamiltonian of monolayer graphene with RSO
interaction is in good approximation equivalent to the Hamiltonian of bilayer graphene\cite{mccann:086805} and
that RSO interaction in monolayer graphene leads to trigonal warping (TW) of the energy bands.
This equivalence of the Hamiltonians may seem surprising since the two physical systems are very different.
We show numerically that as long as RSO interaction is stronger than ISO, it affects the band structure
in two important ways: i) TW changes the topology of the low energy bands at the $\mathbf{K}$ ($\mathbf{K'}$)
points which may be detected as a
\emph{tree-fold increase} in the minimal conductivity of the sample at very low electron densities.
ii) RSO is also manifested through the non-isotropic spin-splitting of the bands
and in particular, through the non-constant band splitting at the $\mathbf{K}$ ($\mathbf{K}'$) points.

\emph{Tight binding model.}---
We start our discussion by considering the tight binding (TB) Hamiltonian of monolayer graphene which
includes RSO coupling\cite{ref:kane-mele-2} as well.
It is given by $H = H_0 + H_{R},$ where
\begin{subequations}
\begin{equation}
\label{eq:H_0}
 H_0 = -\gamma_0\sum\limits_{\left\langle i,j\right\rangle,\sigma}\left({a}^{\dagger}_{i\sigma}
{b}_{j\sigma} + {\rm h.c.}\right).
\end{equation}
Here $\gamma_0$ is the hopping amplitude between nearest neighbor atoms,
${a}^{\dagger}_{i\sigma}$ (${a}_{i\sigma}$)
creates (annihilates) an electron in the $i$th unit cell with
spin $\sigma$ on sublattice $A$, while ${b}^{\dagger}_{j\sigma}$ (${b}_{j\sigma}$)
has the same effect on sublattice $B$.
The sum is taken over nearest neighbor atoms $\left\langle i,j\right\rangle$ and $\textrm{h.c.}$
stands for hermitian conjugate.
The Hamiltonian $H_{R}$ describes the Rashba spin-orbit coupling and it reads\cite{ref:kane-mele-2}:
\begin{equation}
\label{eq:H_RSO}
 H_{R} = i\, \lambda_R\sum\limits_{\left\langle i,j\right\rangle,\mu,\nu}
 \left[{a}^{\dagger}_{i\mu}\left(\boldsymbol{s}_{\mu \nu}\times
\mathbf{\widehat{d}}_{\left\langle i,j\right\rangle} \right)_z {b}_{j\nu} -  h.c. \right]\;,
\end{equation}%
\label{eq:tb-full-ham}
\end{subequations}%
where $\boldsymbol{s}= (s_x,s_y,s_z)$ are the Pauli matrices representing the electron spin operator
and $\mu, \nu =1,2$ denote the $\mu \nu$ matrix elements of the Pauli matrices. 
Moreover, $ \mathbf{\widehat{d}}_{\left\langle i,j\right\rangle} = \mathbf{d}_{\left\langle i,j\right\rangle}/d $
are unit vectors, where $\mathbf{d}_{\left\langle i,j\right\rangle}$
points from atom $j$ to its nearest neighbors $i$ and $d=|\mathbf{d}_{\left\langle i,j\right\rangle}|$.
The strength of the spin-orbit coupling is denoted by $\lambda_R$ which may arise due to a perpendicular
electric field or interaction with the substrate.

Using the operator ${\alpha}^{\dagger}_{\mathbf{q}\sigma} =
\frac{1}{\sqrt{N}}\sum_{i}e^{{\rm i}\mathbf{q}\mathbf{R}_i}{a}^{\dagger}_{i\sigma}$, where $\mathbf{R}_i$
is the   Bravais vector of the $i$th unit cell
and $\textbf{q}$ lies in the first Brillouin zone (and similarly, introducing
${\beta}^{\dagger}_{\mathbf{q}\sigma}$ acting on sublattice $B$) it is easy to find that
\begin{subequations}
\begin{eqnarray}
 H_0 &=& -\gamma_0\sum_{\mathbf{q},\sigma}
 \left[
 f(\mathbf{q})\, {\alpha}^{\dagger}_{\mathbf{q}\sigma}{\beta}_{\mathbf{q}\sigma} + {\rm h.c.}
 \right], \\
 H_{R} &=& i\lambda_R\sum\limits_{\mathbf{q},\mu\nu}\left[{\alpha}^{\dagger}_{\mathbf{q}\mu}
 \left(\boldsymbol{s}_{\mu \nu}\times\mathbf{D}(\mathbf{q}) \right)_z {\beta}_{\mathbf{q}\nu}
 - \textrm{ h.c.} \right] 
\end{eqnarray}%
where
\begin{equation}
f(\mathbf{q}) = \sum_{j=1}^3 e^{-{\rm i}\mathbf{q}\mathbf{a}_j}, \textrm{and }
\mathbf{D}(\mathbf{q}) = - \sum_{j=1}^3 \mathbf{\widehat{d}}_j e^{-{\rm i}\mathbf{q}\mathbf{a}_j}.
\end{equation}
\end{subequations}%
Here $\mathbf{a}_1$, $\mathbf{a}_2$ are the lattice vectors of graphene and it was convenient to introduce
the vector $\textbf{a}_3=0$. Finally, the matrix representation of the Hamiltonian
$H= H_0 +H_R$ acting on wave function ${\Psi}=(\Psi_{A\uparrow},\Psi_{B\uparrow},\Psi_{A\downarrow},
\Psi_{B\downarrow})$ can be written as
\begin{equation}
       H(\mathbf{q}) = \left(
		\begin{smallmatrix}
                  0 & -\gamma_0 f(\mathbf{q}) & 0 & -\lambda_R D_+(\mathbf{q}) \\[1ex]
		-\gamma_0 f^*(\mathbf{q}) & 0 & -\lambda_R D_-^*(\mathbf{q})& 0 \\[1ex]
		0 & -\lambda_R D_-(\mathbf{q}) & 0 & -\gamma_0 f(\mathbf{q}) \\[1ex]
                -\lambda_R D_+^*(\mathbf{q}) & 0 & -\gamma_0 f^*(\mathbf{q}) & 0
                 \end{smallmatrix}
		      \right), 
\label{eq:Hk}
\end{equation}
where  $D_{\pm}(\mathbf{q}) = \pm D_x(\mathbf{q}) - iD_y(\mathbf{q})$. This TB 
Hamiltonian can be used for calculations in the whole Brillouin zone (BZ).
As it has been shown in Ref.~\onlinecite{PhysRevB.79.165442},
the Hamiltonian in Eq.~(\ref{eq:Hk}) preserves the particle-hole symmetry
therefore in the following we will focus on the conduction bands only.
One finds  that the electron and hole bands touch at the $\mathbf{K}$ and
$\mathbf{K'}$ points of the Brillouin zone (here ${\bf K}= (2{\bf b}_2 +{\bf b}_1)/3$ and
${\bf K}^\prime= (2{\bf b}_1 +{\bf b}_2)/3$, where ${\bf b}_1$ and ${\bf b}_2$ are the reciprocal lattice
vectors of the graphene sheet). Next, it will prove  instructive to derive a low energy Hamiltonian
which describes the
excitations in the vicinity of the band-touching points.

\emph{Continuous model.}---
We expand the Hamiltonian (\ref{eq:Hk}) around the Dirac point $\textbf{K}$ of the BZ in terms of
$\mathbf{k}=\textbf{q}-\textbf{K}$ for  $|\mathbf{k}|\approx 0$.
Then one can perform a unitary transformation using the  matrix
$U = \left( \frac{i+1}{2}I_2 + \frac{i-1}{2}\sigma_z\right)\otimes s_z$
(here $I_2$ is a 2 by 2 unit matrix)
and take the inverse Fourier transformation corresponding to the replacement
$\hbar\mathbf{k}\rightarrow \hat{\mathbf{p}} = -i\hbar\boldsymbol{\nabla}$
(here $\hat{\mathbf{p}} = (\hat{p}_x,\hat{p}_y)$ is the momentum operator).
One  obtains finally the following Hamiltonian for monolayer graphene with RSO interaction:
\begin{equation}
 {H}_K = \begin{pmatrix}
                  0 & v_F \hat{p}_- & 0 & v_{\lambda} \hat{p}_+ \\
		v_F \hat{p}_+ & 0 & -3 i\lambda_R  & 0 \\
		0 & 3 i\lambda_R  & 0 & v_F \hat{p}_- \\
                v_{\lambda} \hat{p}_- & 0 & v_F \hat{p}_+ & 0
                 \end{pmatrix}. \label{eq:HB}
\end{equation}
Here $v_F = 3\gamma_0 d/(2\hbar)$, $v_{\lambda} = 3\lambda_R d/(2\hbar)$
and $\hat{p}_{\pm} = \hat{p}_x\pm i \hat{p}_y$.
Apart from a unitary transformation this Hamiltonian is
the same as the Hamiltonian of  bilayer graphene, if the asymmetry between the on-site
energies of the bottom and top layers and the small hopping amplitude $\gamma_4$ can be
neglected\cite{mccann:086805}.
This conclusion may seem surprising since the TB models 
of the two systems look rather different.
As $v_{\lambda} \ll v_F$, when calculating the band structure
the matrix elements $\left(H_K\right)_{14}$ and $\left(H_K\right)_{41}$ are
small compared to the matrix elements which are proportional to $v_F$
and one can question if it is important to keep them.
Indeed,  the theoretical
treatment of RSO in Refs.~\onlinecite{rashbacikk,ref:ingenhoven,ref:gmitra,ref:kuemmeth} is
based on a Hamiltonian in which $\left(H_K\right)_{14}=\left(H_K\right)_{41}=0$.
Note however, that these terms  cause \emph{trigonal warping} of the bands at low energy and they
can therefore be important since TW results in the change of the topology of the energy bands near the Dirac point.
This can be easily seen by recalling the band structure of bilayer graphene with TW\cite{mccann:086805} or
by looking at the plane wave solutions of the Schr\"odinger equation
$H_K\Psi(\mathbf{k})=E(\mathbf{k})\Psi(\mathbf{k})$.
The four eigenvalues of the Hamiltonian (\ref{eq:HB}) as a function of the
wave number $\mathbf{k} = k(\cos\alpha,\sin\alpha)^T$ are given by
\begin{subequations}
\begin{equation}
 E_n^{\pm}(\mathbf{k}) = \pm \hbar v_F\,
\sqrt{\frac{1}{2}\left[k_\lambda^2 + k^2 \left(2+\beta^2\right) + (-1)^n \sqrt{\Upsilon}\right] },
\end{equation}
where $\beta=v_{\lambda}/v_F = \lambda/\gamma$ is the dimensionless strength of the spin-orbit coupling,
$k_\lambda =  2 \beta /d$, $n=1,2$, and
\begin{equation}
\Upsilon = k_\lambda^4 +  2k^2 k_\lambda^2 (2-\beta^2) + k^4 \beta^2 (4+\beta^2) -
 8 k^3 k_\lambda \beta \sin (3 \alpha).
\end{equation}%
\end{subequations}%
\begin{figure}[tb]
\includegraphics[width=8.5cm]{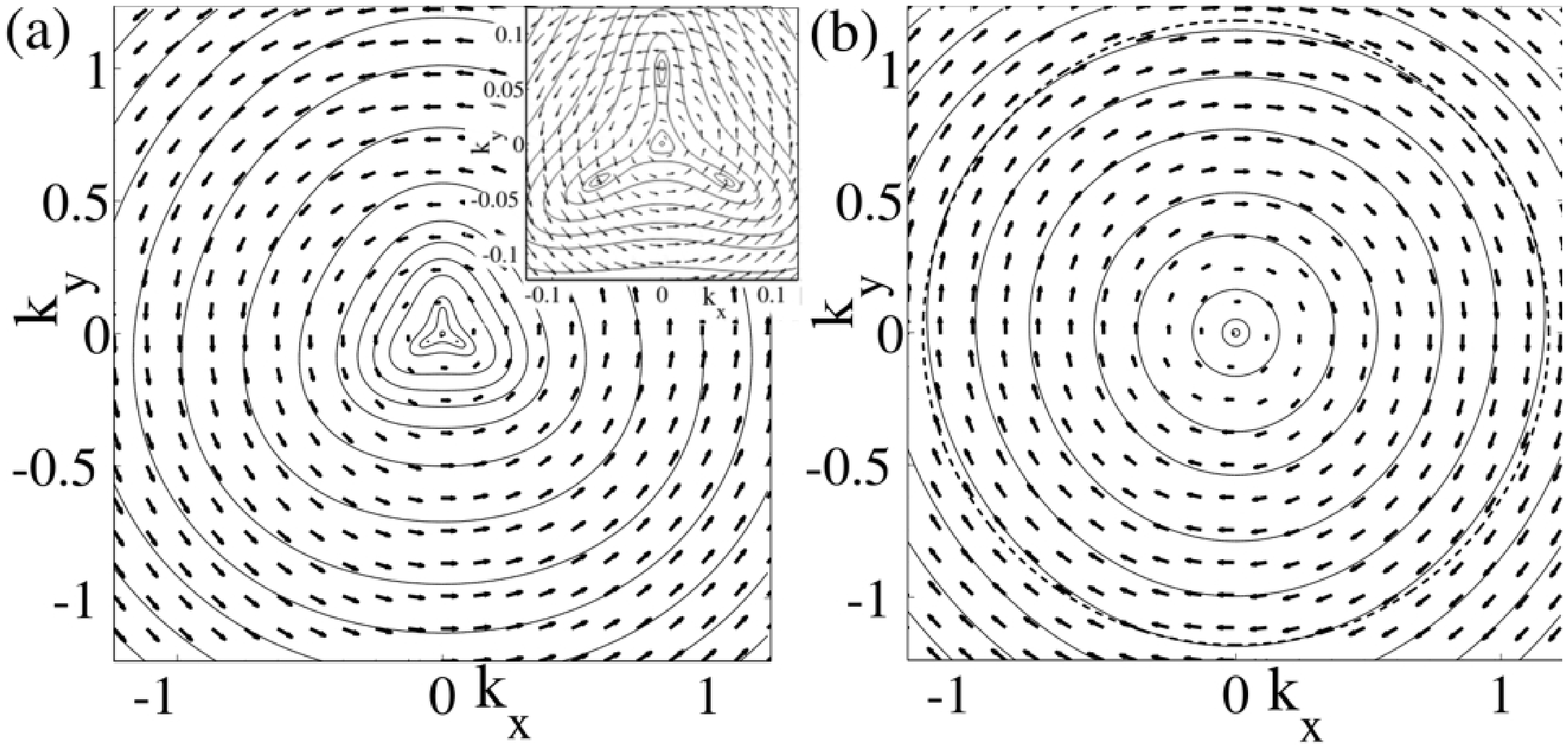}
\includegraphics[width=8.5cm]{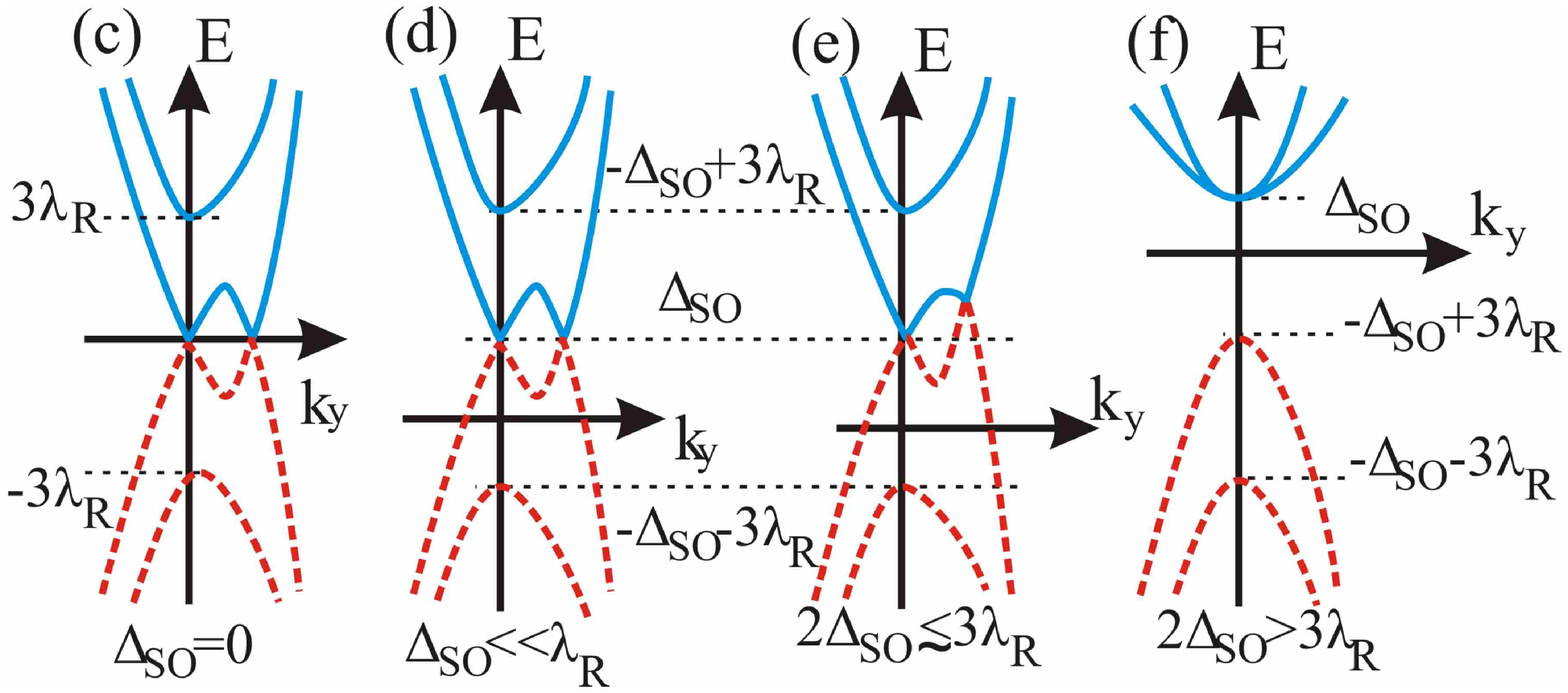}
 \caption{(Color online) The $\mathbf{k}$ dependence and spin polarization of the energy bands: (a)
$E_1^+(\mathbf{k})$ and (b) $E_2^+(\mathbf{k})$ around the $\mathbf{K}$ point
for $\beta=0.034$. Solid lines show the isoenergy contours, the arrows indicate
 the direction of the spin at certain points in the BZ. 
The wave vector components $k_x$, $k_y$ are in units of $3\lambda_R /(2\hbar v_F)$. 
The inset in Fig.~\ref{fig:spectrum}(a) shows a close-up  of 
$E_1^+(\mathbf{k})$ in the vicinity of the $\mathbf{K}$ ($\mathbf{K}^\prime$) point. 
The  pocket structure has a $2\pi/3$ rotational symmetry and
the  distance of the pocket minima from the $\mathbf{K}$ ($\mathbf{K}^\prime$) 
point is $k_{\textrm{SO}}$ (see the text). The dashed line  Fig.~\ref{fig:spectrum}(b) is a 
circle, it indicates that the inosenergy lines of $E_2^+(\mathbf{k})$ are also slightly distorted.
(c)-(f): a cut through the energy bands for $k_x=0$ and (c) $\Delta_{SO}=0$, (d)
$\Delta_{SO}\ll\lambda_R$, (e) $2\Delta_{SO}\lesssim 3\lambda_R$ and (f) $2\Delta_{SO}>3\lambda_R$.
Solid blue lines denote electron bands, while dashed red lines correspond to valance bands.
\label{fig:spectrum}}
\end{figure}
Figures \ref{fig:spectrum}(a) and (b) show the contour plot of the conduction
bands $E_1^+(\mathbf{k})$ and $E_2^+(\mathbf{k})$.
At ${\bf k} = {\bf 0}$ the splitting of the bands is $3\lambda_R$ [see Fig.~\ref{fig:spectrum}(c)].
Similarly to bilayer graphene, the  bands 
have a threefold rotational symmetry.
For  energies smaller than the Lifshitz energy $E_L = 3 \lambda_R \beta^2/(4+\beta^2)$,
the constant energy lines of $E_1^+(\mathbf{k})$ are broken
into four pockets, which can be referred to as one central and three leg parts 
[this is shown in the  inset of Fig.~\ref{fig:spectrum}(a)].
The eigenvalues $\pm E_1^+(\mathbf{k})$ become zero at the $\mathbf{K}$ point of the Brillouin zone, ie, at
$k =  0$, and at the center of the three leg parts located at $k = k_{\textrm{SO}}$ and
$\alpha = -\pi/6; \pi/2; 7\pi/6$.
The distance between the center of the leg parts and the Dirac-point ($\mathbf{K}$ point)
in continuous model is  $k_{\textrm{\textrm{SO}}} = \beta^2\, \frac{2}{d}$,
which is equal to $k_{\textrm{SO}}$ in Ref.~\onlinecite{PhysRevB.79.165442} for the experimentally
relevant case of $\lambda_R \ll \gamma_0$.
The authors of  Ref.~\onlinecite{PhysRevB.79.165442}
have also noticed the three additional zero-energy points
around the  $\mathbf{K}$  point  but it was not recognized that the Hamiltonian (\ref{eq:HB}) of
monolayer graphene with RSO is equivalent to the Hamiltonian of bilayer graphene with TW.
An important difference however between bilayer graphene and monolayer graphene with RSO is that the
splitting $\Delta E(\mathbf{k})=E_2^+(\mathbf{k})-E_1^+(\mathbf{k})$
of the two conduction bands is much smaller in the latter.
This energy is in the range of $0.3-0.48 {\rm eV}$\cite{gamma1param} for bilayer whereas
a recent experiment\cite{PhysRevLett.101.157601} shows that the splitting is about $13{\rm meV}$ for
RSO  in monolayer graphene. Therefore in the latter system even for small electron densities
states in both  $E_1^+(\mathbf{k})$ and  $E_2^+(\mathbf{k})$ would be occupied, whereas in the case
of bilayer graphene the occupation of the upper band can usually be neglected.

So far in our discussion we have completely neglected the intrinsic spin-orbit coupling ISO.
As the Lifshitz energy $E_L$ itself is a very small energy scale,
it is important to understand whether the trigonal warping and the pocket structure survive if a small,
yet finite ISO is also present.
To this end we have numerically calculated
the band structure using the Hamiltonian $H=H_0+H_R+H_{SO}$ with $H_{SO}=\Delta_{SO}\sigma_z s_z$
(see Ref.~\onlinecite{kane-mele})
where $\sigma_z=1$($-1$) describes state on sublattice $A$ ($B$) . The results of the calculations are shown
in Figs.~\ref{fig:spectrum}(c)-(f). 
For $\lambda_R \gg \Delta_{SO}$ the topology of the lower energy band around the  band touching points
is preserved [though the bands get a global shift, compare Fig.~\ref{fig:spectrum}(c) and (d)].
Furthermore, if $2\Delta_{SO}\lesssim 3\lambda_R$, most of the characteristics of the RSO spectrum survive,
but the minima of the leg pockets will be at slightly higher energy than the minimum of the central pocket
[see Fig.~\ref{fig:spectrum}(e)].
Finally, if the ISO splitting become larger than $3\lambda_R$,  the pocket structure vanishes
completely [Fig.~\ref{fig:spectrum}(f)]. Our findings in Figs.~\ref{fig:spectrum}(c)-(d)
are in  good qualitative agreement with the density functional
calculations of Ref.~\onlinecite{ref:gmitra} except that in the results of Ref.~\onlinecite{ref:gmitra}
the TW seems to be absent. This is not surprising however, since for the weak Rashba splitting assumed there
($3\lambda_R \approx 25 \mu{\rm eV}$) trigonal warping is negligible. For stronger $\lambda_R$ values
and if  $\lambda_R \gg \Delta_{SO}$,  which is likely to be the case in the experiment
of Refs.~\onlinecite{PhysRevLett.101.157601,ref:gierz},  TW is expected to be more important.

We now turn to discuss some of the experimental implications of our results and
compare the predictions based on the Hamiltonian (\ref{eq:HB}) and (\ref{eq:Hk})
to some of the results of recent experiments.
One can immediately see that
if the  RSO coupling is much larger than the ISO,
our results predict  that due to the pocket structure shown in the inset 
of Fig.~\ref{fig:spectrum}(a)  for very low electron densities
the minimal conductivity is three times larger than what one would obtain by  neglecting TW.
(For the relevant calculations  for bilayer graphene see Refs.~\onlinecite{cserti:066802,anisotrop}.)

Another way to study the TW experimentally would be to scan the spin splitting of the bands across the BZ.
\begin{figure}[hbt]
 \includegraphics[width=8cm]{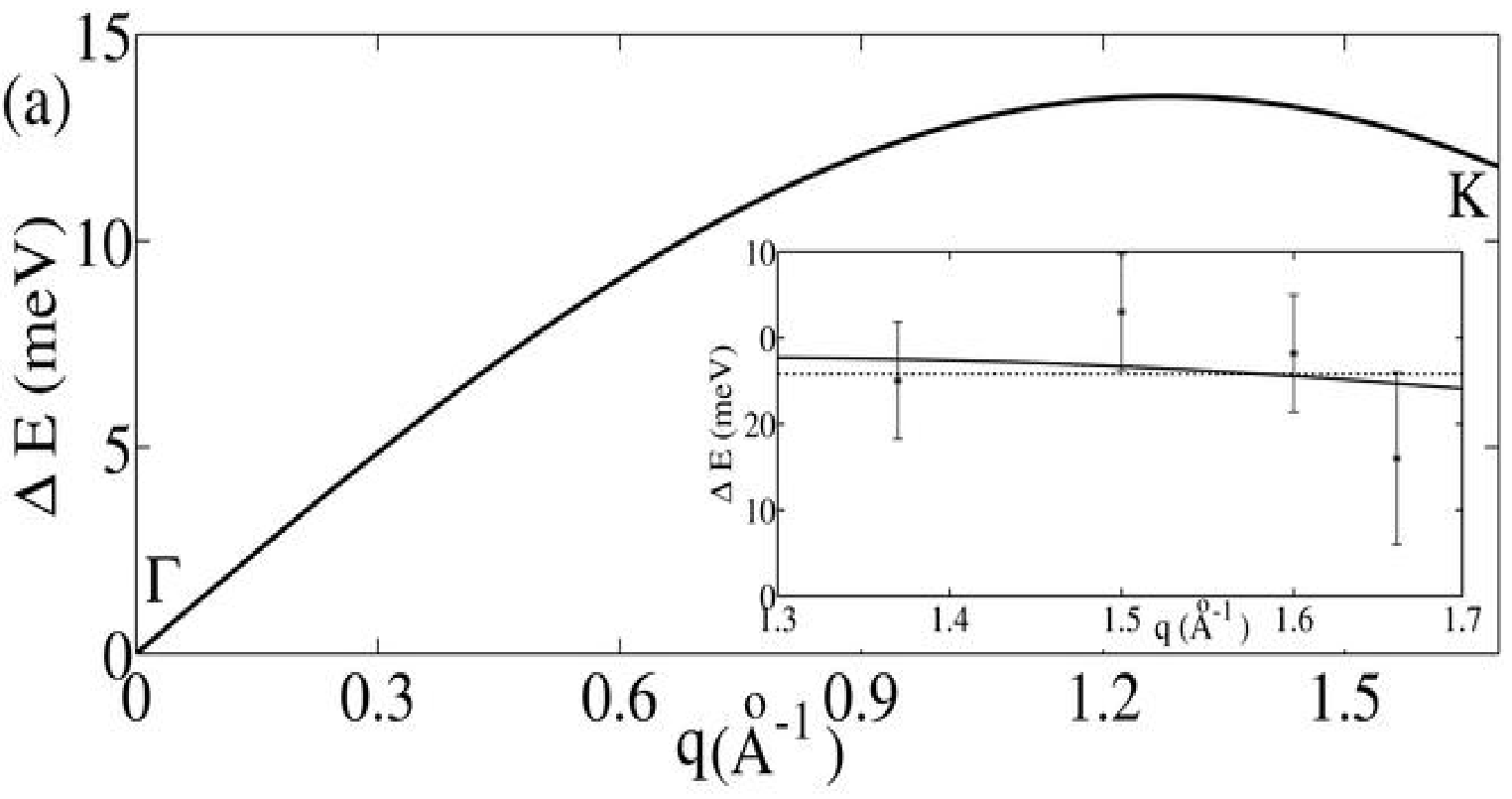}
 \includegraphics[width=8cm]{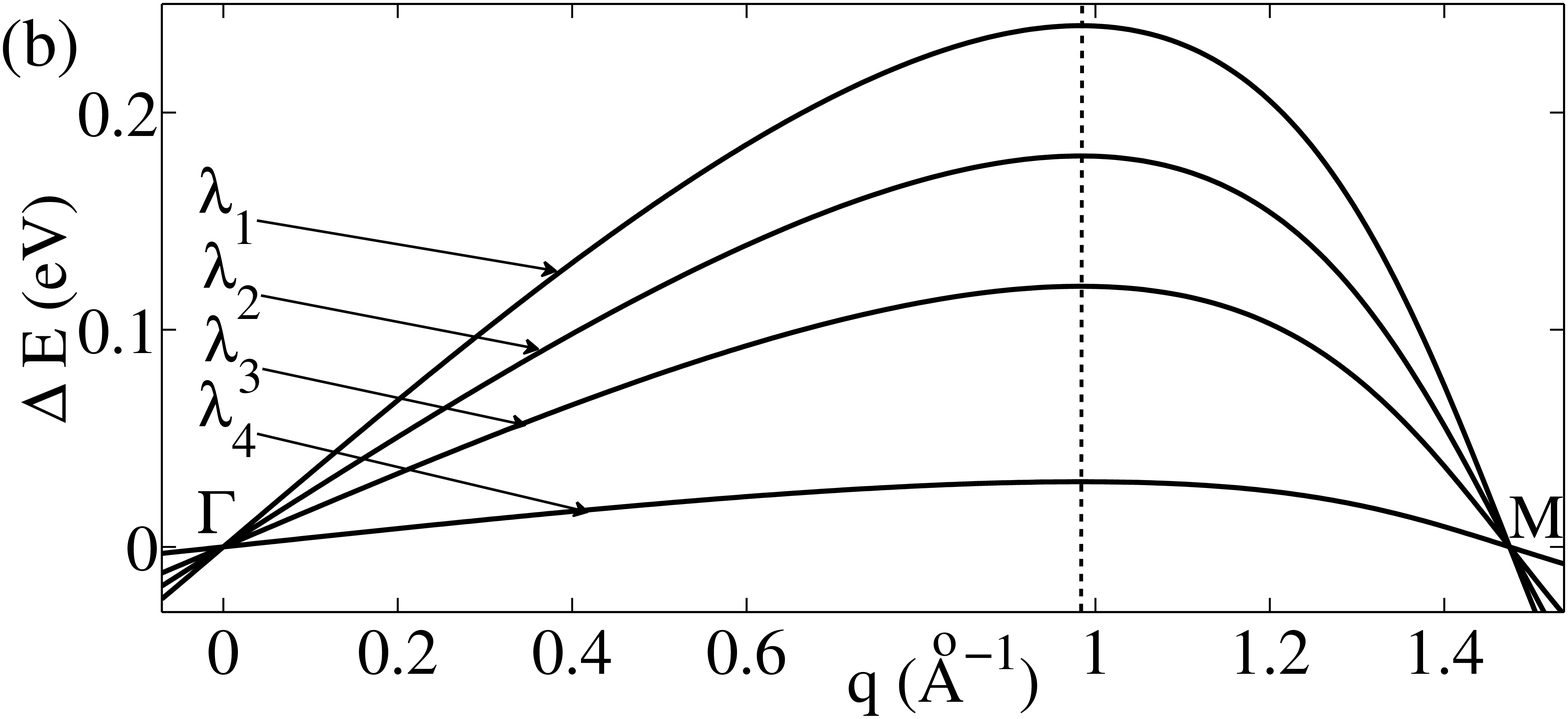}
 \caption{(a) The spin splitting along the $\overline{\Gamma K}$ line of the BZ
 using the Hamiltonian of Eq.~(\ref{eq:Hk}).
The inset shows the comparison  of the experimental data  from Ref.~\onlinecite{PhysRevLett.101.157601}
and the  calculated band splitting with finite TW (solid line) and with no TW (constant dashed line).
(b) The band splitting along $\overline{\Gamma M}$ 
for $\lambda_{R_1}=80\,{\rm meV}$,
$\lambda_{R_2}=60\,{\rm meV}$, $\lambda_{R_3}=40\,{\rm meV}$ and $\lambda_{R_4}=10\,{\rm meV}$
obtained from Hamiltonian (\ref{eq:Hk}). Vertical dashed line indicates the minimum of the curves.
The TB hopping parameter was $\gamma_0=3{\rm eV}$.}
\label{fig:gammaK}
\end{figure}
By looking at Figs.~\ref{fig:spectrum}(a) and (b) it is easy to see that  the splitting
is anisotropic  around the $\mathbf{K}$ point and would reflect a threefold rotational symmetry.
This is an important difference compared to the modell used in
Refs.~\onlinecite{rashbacikk,ref:ingenhoven,ref:gmitra,ref:kuemmeth}.
In contrast to the measurements of Ref.\onlinecite{ref:gierz}, in this model
the splitting does not go to zero along the $\overline{\Gamma K}$ line, which may indicate
a strong influence of the SiC substrate that  the theoretical models considered so far
do not capture. 
Since the Hamiltonian (\ref{eq:HB}) can only be applied around the $\bf K$ point, in order to explore a
larger part of the BZ we also performed calculations using  the  TB model [see Eq.~(\ref{eq:Hk})].
 The band splitting along  the $\overline{\Gamma K}$ line
assuming RSO coupling constant $3\lambda_R=12 {\rm meV}$  
is shown in Fig.~\ref{fig:gammaK}(a).
The most important feature to be observed is that the RSO splitting in the vicinity of the
$\mathbf{K}$ point is not constant as predicted by the model which neglects TW,
but has a small curvature. This can also be seen
in the  inset of Fig.~\ref{fig:gammaK}(a) which shows the comparison of the calculations and of the data
from Fig.~4(c) of Ref.~\onlinecite{PhysRevLett.101.157601}.
The experimental data are consistent with the calculated curve, unfortunately the measurement
resolution does not allow to make a definitive judgment as to whether the splitting is constant or not.
In Fig.~\ref{fig:gammaK}(b) we plot the spin splitting along the $\overline{\Gamma M}$ line in the BZ
for various RSO coupling strengths. The maximum of the band
splitting equals  $3\lambda_R$ and it is always attained at the same $k$ wavenumber, not depending on $\lambda_R$.
It is interesting to note that the shape of these curves  is very reminiscent of the
measurements shown in Fig.3(c) of Ref.~\onlinecite{dedkov}, which were made
on graphene on Ni substrate. The reason of the splitting  in this system, however, appears to
be unclear\cite{Varykhalov2,VarykhalovPRB}.

The SARPRES technique\cite{review} can not only probe the band structure but also the
spin polarization of the bands. The results of Ref.~\onlinecite{PhysRevLett.101.157601} suggest
that the spin polarization has  circular symmetry  around the  $\bf K$ point and
that the spin is polarized in clockwise (anticlockwise) directions in the two bands at the Fermi energy.
Our calculations for the spin polarization of the energy bands $E_1^+(\mathbf{k})$ and $E_2^+(\mathbf{k})$
are in  agreement with this experimental result
[see in  Figs.~\ref{fig:spectrum}(a) and (b), respectively].
It might seem  surprising that the spin polarization  of the band
$E_1^+(\mathbf{k})$ shows a simple rotational symmetry,
without any signs of  TW observed in the spectrum.
As we will demonstrate below, TW affects the spin polarization only in higher orders of $\bf k$.
The calculation of  spin polarization goes along the lines of a similar calculation
in Ref.~\onlinecite{rashbacikk}.
The expectation value of the spin components on sublattice $A$ is given by
${\bf s}^a 
= \hbar/2(\sin\theta_s^a\cos\varphi_s^a, \sin\theta_s^a\sin\varphi_s^a, \cos\theta_s^a)^T$.
We find that in the lowest orders it depends on $k=|{\bf k}|$ as
\begin{subequations}
\begin{equation}
 \tan\theta^a_{n}({\bf k}) = (-1)^{n+1}\left(\frac{k}{\kappa} -  \frac{\sin3\alpha}{2}\frac{k^2d}{\kappa}\right),
 \label{eq:szogeka1}
\end{equation}
and
\begin{equation}
 \varphi^a_n({\bf k}) = \alpha+(-1)^{n+1}\frac{\pi}{2} - \frac{\cos3\alpha}{2}kd, \label{eq:szogeka2}
\end{equation}
\end{subequations}
where the index $n=1,2$ correspond to energy eigenvalues $E_1^{+}({\bf k})$ (or $E_1^{-}({\bf k})$)
and $E_2^{+}({\bf k})$ (or $E_2^{-}({\bf k})$) respectively, and $\kappa=\beta/d$.
Similar calculations for sublattice $B$ yield
 $\theta^b_{n}({\bf k}) =\pi -\theta^a_{n}({\bf k})$ and
$\varphi^b_n({\bf k}) = \varphi^a_n({\bf k})$. 
As one can see  from Eqs.~(\ref{eq:szogeka1})  and (\ref{eq:szogeka2}),  TW does
not  play role  in leading order of $k$.  This explains
the circular symmetry of the spin polarization in Figs.~\ref{fig:spectrum}(a) and (b).
Moreover, the component of the spin perpendicular to the graphene sheet points into
opposite directions on sublattices $A$ and $B$. Therefore,  as in Ref.~\onlinecite{rashbacikk},
we find that
the average spin ${\bf s} = ({\bf s}^a+{\bf s}^b)/2$ is polarized in the
plane of the graphene sheet, it is perpendicular to the wave vector $\bf k$ and its
magnitude grows up from zero (at $k=0$) to  $\hbar/2$  
 further from the $\bf K$ point, see also in Fig.~\ref{fig:spectrum}(a) and (b).

In summary, we have studied the RSO coupling in graphene monolayer.
We have found that in case  both ISO and  RSO  interactions are
present,  as long as the  RSO coupling is stronger, the low energy bands will show
TW in the vicinity of the  $\bf K$ point of the BZ.
Our predictions for the spin splitting of the bands along certain high symmetry directions
in the BZ  are  consistent with the results of a recent experiment\cite{PhysRevLett.101.157601}
which measured the band splitting due to RSO interaction.
Finally, we have demonstrated that the TW of the bands does not affect the spin polarization in
the lowest order of the momentum therefore it shows rotational symmetry around the $\bf K$ point.
These results might be relevant for other novel low dimensional systems where
RSO coupling is important\cite{review}.

\noindent \emph{Acknowledgements:}
We thank A. Varykhalov for helpful discussions and for providing
 the experimental data used in Fig.\ref{fig:gammaK}(a).
P. R. and J. Cs. acknowledge the
support of the Marie Curie ITN project NanoCTM (FP7-PEOPLE-ITN-2008-234970)
and the Hungarian Science Foundation OTKA under the contracts No.~75529 and No.~81492.
A. K. was supported by EPSRC.

\bibliographystyle{prsty}

\end{document}